\documentclass[sigconf, pbalance]{acmart}

\usepackage{subcaption}
\usepackage{url}
\usepackage{graphicx} 
\usepackage{amsmath}
\usepackage{xcolor}
\usepackage{etoolbox}
\usepackage{hyperref}
\makeatletter
\providecommand{\@received}[1]{}
\makeatother
\hypersetup{
    colorlinks=true,
    linkcolor=blue,
    citecolor=blue,
    urlcolor=blue
}

\acmConference[EASE 2026]{The 30th International Conference on Evaluation and Assessment in Software Engineering}{9–12 June, 2026}{Glasgow, Scotland, United Kingdom}

\AtBeginDocument{%
  }

\setcopyright{acmlicensed}
\copyrightyear{2018}
\acmYear{2018}
\acmDOI{XXXXXXX.XXXXXXX}
\acmISBN{978-1-4503-XXXX-X/2018/06}

\begin{document}

%%
%% The "title" command has an optional parameter,
%% allowing the author to define a "short title" to be used in page headers.
\title{Sustainability Analysis of Prompt Strategies for SLM-based Automated Test Generation}

%%
%% The "author" command and its associated commands are used to define
%% the authors and their affiliations.
%% Of note is the shared affiliation of the first two authors, and the
%% "authornote" and "authornotemark" commands
%% used to denote shared contribution to the research.
% \author{Ben Trovato}
% \authornote{Both authors contributed equally to this research.}

% \orcid{1234-5678-9012}
\author{Pragati Kumari}
\affiliation{%
  \institution{University Of Calgary}
  \city{Calgary}
  \country{Canada}}
\email{pragati.kumari@ucalgary.ca}

\author{Novarun Deb}
\affiliation{%
  \institution{University Of Calgary}
  \city{Calgary}
  \country{Canada}}
\email{novarun.deb@ucalgary.ca}

% \author{Valerie B\'eranger}
% \affiliation{%
%   \institution{Inria Paris-Rocquencourt}
%   \city{Rocquencourt}
%   \country{France}
% }

% \author{Aparna Patel}
% \affiliation{%
%  \institution{Rajiv Gandhi University}
%  \city{Doimukh}
%  \state{Arunachal Pradesh}
%  \country{India}}

% \author{Huifen Chan}
% \affiliation{%
%   \institution{Tsinghua University}
%   \city{Haidian Qu}
%   \state{Beijing Shi}
%   \country{China}}

% \author{Charles Palmer}
% \affiliation{%
%   \institution{Palmer Research Laboratories}
%   \city{San Antonio}
%   \state{Texas}
%   \country{USA}}
% \email{cpalmer@prl.com}

% \author{John Smith}
% \affiliation{%
%   \institution{The Th{\o}rv{\"a}ld Group}
%   \city{Hekla}
%   \country{Iceland}}
% \email{jsmith@affiliation.org}

% \author{Julius P. Kumquat}
% \affiliation{%
%   \institution{The Kumquat Consortium}
%   \city{New York}
%   \country{USA}}
% \email{jpkumquat@consortium.net}

%%
%% By default, the full list of authors will be used in the page
%% headers. Often, this list is too long, and will overlap
%% other information printed in the page headers. This command allows
%% the author to define a more concise list
%% of authors' names for this purpose.
\renewcommand{\shortauthors}{Kumari and Deb}

%%
%% The abstract is a short summary of the work to be presented in the
%% article.
\begin{abstract}
The growing adoption of prompt-based automation in software testing raises important issues regarding its computational and environmental sustainability. Existing sustainability studies in AI-driven testing primarily focus on large language models, leaving the impact of prompt engineering strategies largely unexplored—particularly in the context of Small Language Models (SLMs). This gap is critical, as prompt design directly influences inference behavior, execution cost, and resource utilization, even when model size is fixed.
To the best of our knowledge, this paper presents the first systematic sustainability evaluation of prompt engineering strategies for automated test generation using SLMs. We analyze seven prompt strategies across three open-source SLMs under a controlled experimental setup. Our evaluation jointly considers execution time, token usage, energy consumption, carbon emissions, and {\color{black} coverage} test quality, the latter assessed through coverage analysis of the generated test scripts.
The results show that prompt strategies have a substantial and independent impact on sustainability outcomes, often outweighing the effect of model choice. Reasoning-intensive strategies such as Chain of Thought and Self-Consistency achieve higher coverage but incur significantly higher execution time, energy consumption, and carbon emissions. In contrast, simpler strategies such as Zero-Shot and ReAct deliver competitive {\color{black} coverage} test quality with markedly lower environmental cost, while Least-to-Most and Program of Thought offer balanced trade-offs.
\end{abstract}

%%
%% The code below is generated by the tool at http://dl.acm.org/ccs.cfm.
%% Please copy and paste the code instead of the example below.
%%

%%
%% Keywords. The author(s) should pick words that accurately describe
%% the work being presented. Separate the keywords with commas.
%% A "teaser" image appears between the author and affiliation
%% information and the body of the document, and typically spans the
%% page.
% \begin{teaserfigure}
%   \includegraphics[width=\textwidth]{sampleteaser}
%   \caption{Seattle Mariners at Spring Training, 2010.}
%   \Description{Enjoying the baseball game from the third-base
%   seats. Ichiro Suzuki preparing to bat.}
%   \label{fig:teaser}
% \end{teaserfigure}

%\received{20 February 2007}
% \received[revised]{12 March 2009}
% \received[accepted]{5 June 2009}

%%
%% This command processes the author and affiliation and title
%% information and builds the first part of the formatted document.

\settopmatter{printacmref=false} % Removes ACM reference format
\setcopyright{none}               % Removes copyright box
\renewcommand\footnotetextcopyrightpermission[1]{} % Removes footnote with conference info
\pagestyle{plain}                 % Removes running headers

\maketitle

\section{Introduction}\label{sec:sec1}
Automated testing frameworks such as JUnit and Selenium, together with continuous integration and delivery pipelines, enable tests to be executed repeatedly across development cycles. While this automation improves software quality and development productivity, frequent test execution introduces non-trivial computational and energy costs, motivating increasing interest in sustainability-aware testing practices \cite{verdecchia_energy_testing, zaidman_ast2024}.

Recent advances in large and small language models have transformed software testing by enabling automated unit test generation and augmentation. Empirical studies demonstrate that language models can generate useful unit tests under realistic conditions, although effectiveness varies across tasks, prompt designs, and execution settings \cite{schaefer2024_llm_unit_tests}. Hybrid approaches combining search-based testing with language model guidance show that interaction patterns and prompt structures can significantly influence coverage outcomes \cite{lemieux2023codamosa}. At the same time, survey studies indicate that prompt engineering is pervasive in language model based testing workflows, while evaluation practices remain heterogeneous \cite{wang2024_llm_testing_survey}. Notwithstanding these advances, designing, validating, and refining correct and runnable test scripts continues to require substantial human effort.

Despite growing evidence of effectiveness, the environmental impact of language model assisted testing remains less systematically studied. Recent work shows that even small language models can incur substantial energy overheads when used for automated unit test generation, underscoring the need to treat sustainability as a first-class evaluation dimension in testing workflows \cite{durelli2025_energy_footprint_slm_tests}. More broadly, emerging research on efficient and green language models for software engineering emphasizes the importance of balancing quality improvements with computational, energy, and carbon costs \cite{shi2025_green_llm4se}. However, task-specific analyses that jointly examine prompt strategies, sustainability metrics, and testing effectiveness are still limited.

This paper investigates how prompt strategy selection influences sustainability and {\color{black} coverage} test quality in automated unit test generation using small language models. We analyze three open-source SLMs deployed with 4-bit quantization to reflect resource-constrained inference settings commonly encountered in practice. Across these models, we systematically evaluate seven prompt strategies: Zero-shot \cite{cheng2025revisiting,dellaporta2025promptpatterns}, Few-shot\cite{tang2025fewshot}, Chain-of-Thought\cite{wei2022chain}, Least-to-Most\cite{zhou2023least}, Program-of-Thought\cite{payoungkhamdee2025potmultilingual}, Self-Consistency\cite{wang2023selfconsistency,pan2025modularization}, and ReAct\cite{react2023}. For each model and prompt combination, we quantify total token usage (by aggregating input and output tokens), execution duration, CPU, GPU, and RAM energy consumption, and carbon emissions measured using CodeCarbon\footnote{\url{https://codecarbon.io/}}. Test effectiveness is assessed using normalized code coverage computed via \texttt{coverage.py}.

The results show that prompt strategy selection can substantially influence tokenization behavior, runtime, energy consumption, and carbon emissions, even when the underlying model and task remain unchanged. Reasoning-intensive prompt strategies often improve coverage but incur disproportionately higher sustainability costs, whereas simpler or more structured prompts exhibit more favorable {\color{black} coverage} quality–cost trade-offs. These observations highlight prompt design as an important and actionable factor in sustainability-aware software testing.

\textbf{Contributions.} The main contributions of this paper are as follows:
(I) We present a prompt-centric sustainability evaluation framework for automated unit test generation using small language models, integrating token usage, execution time, energy consumption, carbon emissions, and coverage test quality.
(II) We apply this framework to carry out a controlled experimental evaluation of seven prompt strategies across three quantized open-source SLMs, enabling a systematic comparison of sustainability and {\color{black} coverage} quality trade-offs.
(III) We analyze the experimental results to examine how different prompt strategies influence execution behavior and sustainability-related metrics in automated testing.

The remainder of this paper is organized as follows. Section \ref{sec:sec2} presents the related work. Section \ref{sec:sec3} describes the proposed framework. Section \ref{sec:sec4} details the experimental setup. Section \ref{sec:sec5} reports the results. Section \ref{sec:sec6}  discusses the findings. Section \ref{sec:sec7}  outlines threats to validity. Finally, Section \ref{sec:sec8}  concludes the paper and discusses directions for future work.

\section{Related Work}\label{sec:sec2}

Recent research has highlighted prompt engineering as a promising lever for reducing the environmental cost of language model inference, particularly in code generation and testing tasks. Prior studies systematically evaluate zero-shot, one-shot, and few-shot prompting variants, as well as custom prompt tags, using benchmarks such as CodeXGLUE \cite{lu2021codexglue} and locally deployed models. These works measure execution time and energy consumption using tools like CodeCarbon and assess output quality via exact match and edit distance, demonstrating that carefully designed prompt structures can significantly reduce GPU energy usage without degrading generation quality \cite{rubei2025prompt}. Complementary analyses further reveal that non-semantic prompt characteristics, including code formatting elements such as indentation, whitespace, and newlines, can substantially inflate token counts and computational cost, motivating format-aware prompt optimization strategies \cite{hidden_cost_readability}.

Building on prompt-structure analysis, recent work has examined how prompt linguistic complexity and interaction patterns influence energy--performance trade-offs in software engineering tasks. Green Prompt Engineering explicitly frames prompt design as a sustainability lever alongside hardware and model selection, showing that simpler, more readable prompts significantly reduce energy and carbon consumption with minimal loss in task performance \cite{green_prompt_engineering}. Related profiling studies using fine-grained monitoring frameworks demonstrate that inference duration and response token length are strongly correlated with energy usage, while prompt linguistic complexity alone plays a comparatively smaller role \cite{price_of_prompting}. Empirical studies on automated unit test generation further show that prompt design and interaction structure can substantially influence test effectiveness and coverage outcomes \cite{schaefer2024_llm_unit_tests, lemieux2023codamosa}. However, these works primarily emphasize effectiveness or general inference behavior rather than sustainability-aware evaluation.

Sustainability-oriented research, at a broader systems level, has analyzed the life-cycle energy and carbon footprint of AI-driven systems, demonstrating that inference-time usage and large-scale deployment can rival or exceed training-related emissions \cite{llm_lifecycle_energy}.
In parallel, Green IT and Green Software research establishes foundational principles for reducing the environmental impact of software systems, emphasizing that software design and execution decisions significantly influence long-term energy usage across large-scale infrastructures \cite{green_it_software_2021}. While these works provide high-level sustainability guidance, they do not examine prompt-level decisions or inference-time trade-offs in AI-driven workflows.

\begin{figure*}[!ht]
    \centering
    \includegraphics[width=0.7\textwidth]{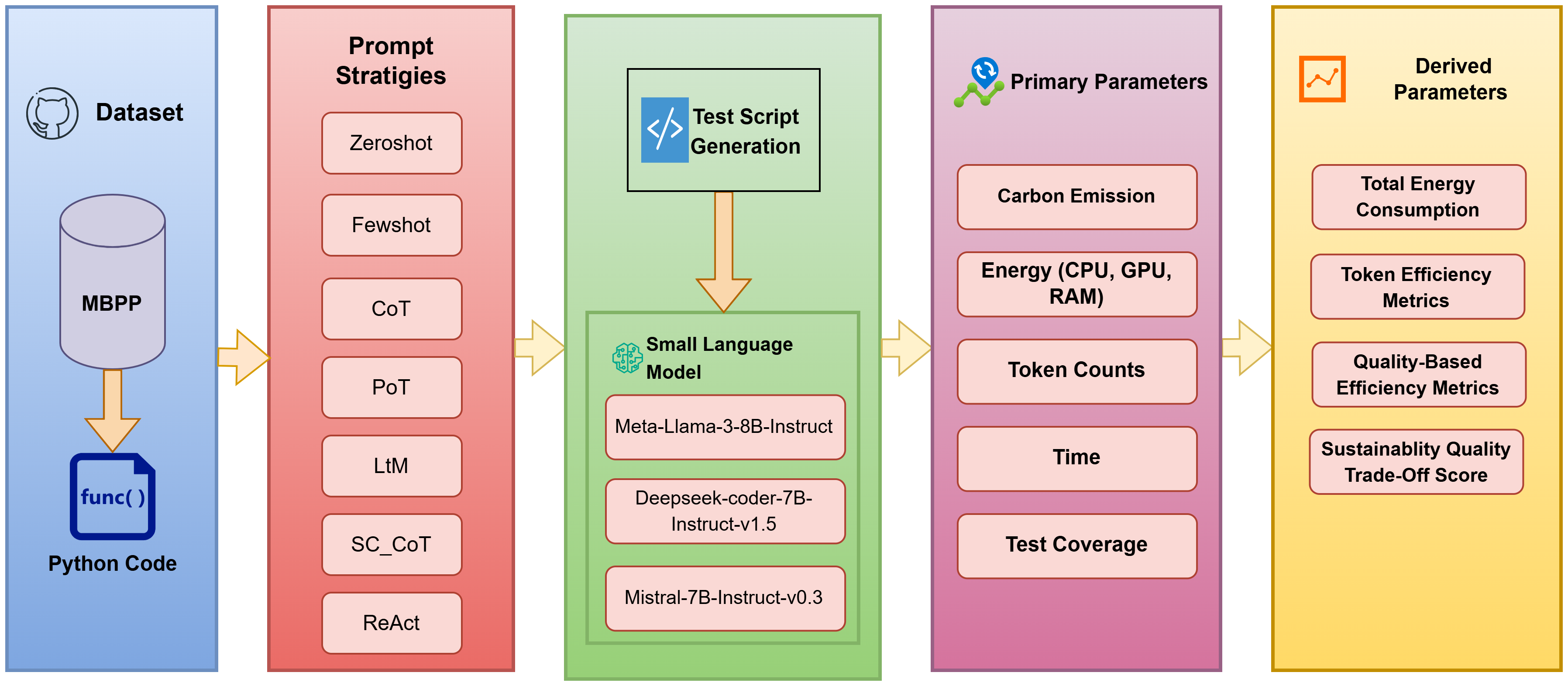}
    \caption{The experiment framework.}
    \label{fig:workflow}
\end{figure*}

Recent work has also examined the role of AI-assisted development tools in improving software productivity and sustainability from a human-centered perspective. Using mixed-method approaches and large-scale surveys, these studies report benefits in terms of reduced development time, improved testing efficiency, and evolving prompt usage practices within software engineering workflows \cite{ai_tools_sustainability_ssrn, wang2024_llm_testing_survey}. However, such perspectives focus on perceived productivity and adoption trends rather than quantifying energy consumption, carbon emissions, or inference-time costs introduced by prompt design.

Energy-aware software testing has emerged within software engineering as a response to the overlooked environmental impact of testing and quality assurance activities.
Foundational work introduces energy-aware testing as a paradigm that treats energy consumption as a first-class concern, proposing strategies such as energy-aware test prioritization, batching, and execution scheduling to reduce testing energy usage without significantly impacting effectiveness \cite{verdecchia_energy_testing}. Empirical studies further quantify the electricity consumption of automated testing and continuous integration pipelines, revealing substantial variation in energy usage across projects and highlighting the non-trivial cost of frequent test execution \cite{zaidman_ast2024}. Recent analyses also demonstrate that automated unit test generation using small language models can introduce measurable energy overheads, even under resource-constrained settings \cite{durelli2025_energy_footprint_slm_tests}. Complementary approaches leverage runtime measurements and system call traces to estimate software energy consumption at scale, enabling automated detection of energy regressions during testing \cite{greenoracle}. While these efforts advance sustainability-aware testing methodologies, they do not address prompt-level optimization in AI-driven test generation pipelines.

Recent advances in prompt-based reasoning have explored structured prompting strategies such as chain-of-thought, least-to-most, and action-based prompting to improve reasoning accuracy in language models. ReAct introduces a unified prompting paradigm that interleaves reasoning traces with task-specific actions, enabling models to dynamically plan, retrieve information, and adapt decisions during execution \cite{react2023}. Parallel work on green and efficient language models for software engineering emphasizes balancing quality improvements with computational, energy, and carbon costs across tasks \cite{shi2025_green_llm4se}. These approaches demonstrate that prompt structure significantly influences model behavior and execution dynamics, making them directly relevant to prompt-strategy-driven sustainability evaluation frameworks.

 Existing literature establishes prompt engineering as an important factor in inference efficiency, highlights energy-aware testing as a necessary response to sustainable software development, and underscores the growing environmental impact of AI-driven systems. However, current studies either focus on general inference behavior, prompt readability, reasoning accuracy, or process-level sustainability, without jointly analyzing prompt strategies, environmental cost, and {\color{black} coverage} test quality in automated software testing. This gap motivates our work, which provides a systematic, prompt-centric sustainability evaluation for automated test generation using Small Language Models, explicitly quantifying the trade-offs between energy consumption, carbon emissions, execution efficiency, and test coverage.

\section{Sustainability Analysis of Prompts}\label{sec:sec3}

The overall architecture of the proposed framework is illustrated in Fig.~\ref{fig:workflow}. It is designed to systematically analyze how different prompt strategies influence the sustainability and {\color{black} coverage} quality of automated software testing using Small Language Models (SLMs). The framework is organized as a linear pipeline consisting of five key components, each of which is described below.
\\\\
\noindent\textbf{A. Dataset:} 
The framework operates on the MBPP\footnote{\url{https://github.com/google-research/google-research/tree/master/mbpp}} (Mostly Basic Python Problems) dataset, which consists of a collection of Python programming tasks designed to be solvable by entry-level programmers. Each task represents a standalone unit of executable Python code and serves as the system under test. The dataset provides a structured schema including a unique task identifier, a natural language problem description, a reference code solution, and associated test specifications. Specifically, each task includes a list of automated test cases (\texttt{test\_list}), optional test setup code (\texttt{test\_setup\_code}), and an additional set of challenge or edge-case tests (\texttt{challenge\_test\_list}) when available. This structure enables consistent generation, execution, and evaluation of test scripts across different prompt strategies and model executions, while ensuring a controlled and standardized experimental setting.
\\\\
\noindent\textbf{B. Prompt Strategies:} The prompt strategies component defines how instructions are formulated and provided to the language models for test generation. The framework evaluates seven prompt strategies, including Zeroshot, Fewshot, Chain-of-Thought (CoT), Least-to-Most (LtM), Self-Consistency CoT (SC\_CoT), and ReAct. Each strategy introduces varying levels of structure, reasoning guidance, or interaction patterns, allowing the framework to isolate the impact of prompt design on both computational cost and {\color{black} coverage} test quality.
\\\\
\noindent\textbf{C. Test Script Generation:} 
The generated prompts are executed using Small Language Models, specifically \texttt{Meta-Llama-3-8B-\\-Instruct}, \texttt{DeepSeek-Coder-7B-Instruct-v1.5}, and \texttt{Mistral-\\-7B-Instruct-v0.3}. These models are selected to represent compact, instruction-tuned SLMs commonly used in automated code-related tasks. By keeping model configurations fixed, the framework ensures that observed variations in sustainability and {\color{black} coverage} quality metrics can be primarily attributed to prompt strategies rather than model-side differences. In this stage, the prepared prompts are used to generate automated test scripts. The prompt strategies are applied to the input Python code, guiding the model to synthesize executable unit tests. This component represents the core generation phase of the framework, where prompt design directly influences the nature, length, and structure of the generated test scripts.
\\\\
\noindent\textbf{D. Primary Parameters:} Following test generation, the framework collects primary execution parameters directly from the inference process. These include carbon emissions, energy consumption across CPU, GPU, and RAM, token counts, execution time, and test coverage. These parameters represent raw measurements obtained during execution and form the foundation for subsequent analysis.
\\\\
\noindent\textbf{E. Derived Parameters:} The final component of the framework computes derived parameters from the collected primary measurements. These derived metrics include total energy consumption, token efficiency measures, {\color{black}coverage} quality efficiency metrics, and a sustainability--{\color{black}coverage} quality trade-off score. By transforming raw measurements into normalized and comparative indicators, this component enables systematic evaluation of sustainability and {\color{black}coverage} quality trade-offs across prompt strategies and models.

\subsection{Primary Parameters}

Each model prompt strategy combination is evaluated by executing the automated test generation pipeline and collecting the following primary parameters:

\begin{itemize}
    \item[] -- $T$: Total number of generated tokens, including both input and output tokens  
    \item[] -- $\tau$: Total execution duration measured in seconds  
    \item[] -- $CO_2$: Total carbon emissions measured in kilograms of CO$_2$ equivalent  
    \item[] -- $E_{cpu}$: Energy consumption attributed to CPU usage (kWh)  
    \item[] -- $E_{gpu}$: Energy consumption attributed to GPU usage (kWh)  
    \item[] -- $E_{ram}$: Energy consumption attributed to memory usage (kWh)  
    \item[] -- $Q$: Normalized test coverage
\end{itemize}

All primary parameters are recorded after each model completion. Execution duration, energy consumption, and carbon emissions are measured using CodeCarbon, which provides fine-grained accounting across CPU, GPU, and RAM resources. Test coverage is computed using \texttt{coverage.py} and initially reported as a percentage. To ensure compatibility with subsequent efficiency and sustainability metrics, coverage is normalized to the range $[0,1]$ as:
\begin{equation}
Q = \frac{\text{Coverage}}{100}.
\end{equation}

\textit{Interpretation.}  
$Q$ represents the normalized proportion of program logic exercised by the generated test suite.

\textit{Relevance to sustainability.}  
Normalization enables coverage to be consistently combined with time, energy, and carbon emission metrics in composite sustainability and efficiency measures.\\

\noindent\textbf{Note:}
{\color{black}
\textit{Justification of Coverage as a Quality Proxy.}
While $Q$ is a widely used structural metric in automated testing, it does not capture all dimensions of test quality, such as semantic correctness, fault detection capability, or oracle strength. Therefore, in this study, coverage is treated as a proxy for \emph{structural test adequacy} rather than a comprehensive measure of overall test quality. Accordingly, all quality-related interpretations and conclusions are scoped to coverage evaluation.
}

\subsection{Derived Parameters}

Aggregate energy and normalized time measures are derived from the primary measurement parameters and used in subsequent efficiency and trade-off analyses.

\subsubsection{Total Energy Consumption}

\begin{equation}
E_{tot} = E_{cpu} + E_{gpu} + E_{ram}
\end{equation}

\textit{Interpretation.}  
$E_{tot}$ represents the complete computational energy footprint of a prompt strategy, aggregated across CPU, GPU, and memory usage.

\textit{Relevance to sustainability.}  
Total energy consumption directly reflects operational cost, as higher energy usage corresponds to increased computational resource expenditure during test generation. It also reflects environmental impact because computational energy use is directly associated with carbon emissions, making it a central metric for sustainability evaluation.

\subsubsection{Execution Time in Hours}

\begin{equation}
\tau_{hr} = \frac{\tau}{3600}
\end{equation}

\textit{Interpretation.}  
$\tau_{hr}$ expresses execution time in hours, ensuring consistency with energy and carbon accounting units.

\textit{Relevance to sustainability.}  
Longer execution times typically lead to increased energy consumption and higher carbon emissions during inference.

% \textit{Overall interpretation.}  
% Together, these derived metrics provide a unified representation of computational effort in terms of energy and time, enabling consistent integration into normalized cost, quality efficiency, and sustainability--quality trade-off measures.

\subsubsection{Token Throughput}

\begin{equation}
\text{TokRate} = \frac{T}{\tau_{hr}}
\end{equation}

\textit{Interpretation.} Token throughput measures tokens generated per hour.

\textit{Relevance to sustainability.} Higher throughput indicates more efficient use of compute resources.

\subsection{Cost per 1K Tokens}

Per-1K token metrics are computed to normalize execution cost by output volume, capturing time, carbon emissions, and energy consumption relative to the amount of generated output. These metrics enable fair comparison across prompt strategies with differing token lengths.

% \begin{table}[t]
% \centering
% \caption{Summary of primary and derived metrics}
% \begin{tabular}{lll}
% \toprule
% Metric & Symbol & Purpose \\
% \midrule
% Total tokens & $T$ & Output scale \\
% Duration (sec) & $\tau$ & Execution cost \\
% Carbon emission & $CO_2$ & Environmental impact \\
% Total energy & $E_{tot}$ & Computational cost \\
% Coverage & $Q$ & Test quality \\
% Token throughput & TokRate & Output efficiency \\
% Energy per 1K tokens & EPer1KTok & Energy normalization \\
% Quality per kWh & QPerkWh & Quality efficiency \\
% SQ score & SQScore & Sustainability trade-off \\
% \bottomrule
% \end{tabular}
% \end{table}

\subsubsection{Time per 1K Tokens}

\begin{equation}
\text{SecPer1KTok} = \frac{1000 \cdot \tau}{T}
\end{equation}

\textit{Interpretation.}  This metric represents the average execution time required to generate 1,000 tokens. Lower values indicate faster token generation and higher inference efficiency.

\textit{Relevance to sustainability.} Reduced execution time directly contributes to lower energy consumption and resource utilization, particularly in repeated or large-scale testing workflows.

\subsubsection{Carbon per 1K Tokens}

\begin{equation}
\text{$CO_2$Per1KTok} = \frac{1000 \cdot CO_2}{T}
\end{equation}

\textit{Interpretation.}  
This metric quantifies the amount of carbon emissions associated with producing 1,000 tokens during model inference.

\textit{Relevance to sustainability.}  
This measure enables direct comparison of the carbon efficiency of different prompt strategies by normalizing emissions to output volume, independent of total generation scale.

\subsubsection{Energy per 1K Tokens}

\begin{equation}
\text{EPer1KTok} = \frac{1000 \cdot E_{tot}}{T}
\end{equation}

\textit{Interpretation.}  
This metric captures the total computational energy consumed, across CPU, GPU, and RAM, to generate 1,000 tokens.

\textit{Relevance to sustainability.}  
Lower energy per 1K tokens indicates more energy-efficient inference behavior, which is critical for sustainable deployment of language model driven testing pipelines.

\textit{Overall interpretation.}  
Together, these per-1K token metrics provide a scale-independent view of execution efficiency, enabling systematic comparison of prompt strategies in terms of time, energy, and carbon cost per unit output.

\subsection{{\color{black}Coverage} Quality Efficiency Metrics}

{\color{black}coverage} Quality-normalized efficiency metrics are defined to assess how effectively computational and environmental resources translate into test coverage, relating achieved coverage to token usage, energy consumption, and carbon emissions.

\subsubsection{{\color{black}Coverage} Quality per 1K Tokens}

\begin{equation}
\text{QPer1KTok} = \frac{1000 \cdot Q}{T}
\end{equation}

\textit{Interpretation.}  This metric represents the amount of test coverage achieved for every 1,000 generated tokens, capturing how efficiently generated output contributes to testing {\color{black}coverage} quality.

\textit{Relevance to sustainability.} Higher values indicate prompt strategies that achieve greater coverage with fewer tokens, reducing unnecessary computation and associated energy and carbon costs.

\subsubsection{{\color{black}Coverage} Quality per kWh}

\begin{equation}
\text{QPerkWh} = \frac{Q}{E_{tot}}
\end{equation}

\textit{Interpretation.}  
This metric quantifies the amount of test coverage obtained per unit of total energy consumed during inference.

\textit{Relevance to sustainability.}  
It directly reflects energy efficiency, favoring prompt strategies that maximize testing effectiveness while minimizing energy usage across CPU, GPU, and memory resources.

\subsubsection{{\color{black}Coverage} Quality per kg CO$_2$}

\begin{equation}
\text{QPer$CO_2$} = \frac{Q}{CO_2}
\end{equation}

\textit{Interpretation.}  
This metric measures how much test coverage is achieved per kilogram of carbon emissions produced during model inference.

\textit{Relevance to sustainability.}  
This metric, by normalizing {\color{black}coverage} quality against carbon emissions, highlights prompt strategies that deliver higher testing benefits with lower environmental impact.

\textit{Overall interpretation.}  
Collectively, these {\color{black}coverage} quality efficiency metrics capture how effectively prompt strategies convert computational and environmental resources into testing value, enabling principled comparison of sustainability--{\color{black}coverage} quality trade-offs across models and prompts.

\subsection{Sustainability--{\color{black}Coverage} Quality Trade-off Score}

A composite score is defined to jointly capture testing {\color{black}coverage} quality and sustainability dimensions by integrating test coverage, energy consumption, carbon emissions, and execution time into a single metric.
{\color{black}The formulation is motivated by the need to evaluate how effectively testing outcomes are achieved relative to the computational and environmental cost incurred during inference. By placing coverage ($Q$) in the numerator and cost-related factors in the denominator, the score reflects the efficiency with which testing outcomes are obtained per unit of resource usage.}

This metric enables principled comparison of prompt strategies under different sustainability preferences by combining {\color{black}coverage} quality, energy, emissions, and runtime into a single formulation, without relying on isolated efficiency measures.

\begin{equation}
\text{SQScore} =
\alpha \cdot
\frac{Q}{CO_2 \cdot E_{tot} \cdot \tau_{hr}}
\end{equation}

{\color{black}The multiplicative combination of $CO_2$, $E_{tot}$, and $\tau_{hr}$ is used to capture the joint effect of environmental impact and computational cost, ensuring that increases in any one factor proportionally reduce the overall score. This design emphasizes balanced optimization rather than improvement along a single dimension.}
The weighting factor $\alpha$ controls the relative emphasis between {\color{black}coverage} quality and sustainability objectives and is evaluated under three prioritization settings: $\alpha = 0.5$ (eco-first prioritization), $\alpha = 1.0$ (balanced prioritization), and $\alpha = 2.0$ ({\color{black}coverage} quality-first prioritization).
{\color{black}These settings enable a simple sensitivity analysis by varying the relative importance of coverage versus sustainability cost, allowing comparison of prompt strategies under different practical priorities.}

\textit{Interpretation.}  
The Sustainability--{\color{black}Coverage} Quality Trade-off Score increases when higher test coverage is achieved with lower energy consumption, reduced carbon emissions, and shorter execution time, reflecting more efficient use of computational resources.

\textit{Relevance to sustainability.}  
A single formulation combining {\color{black}coverage} quality, energy, emissions, and runtime enables principled comparison of prompt strategies under different sustainability preferences without relying on isolated efficiency measures. {\color{black}It also provides a normalized basis for comparing trade-offs across models and prompt strategies with differing execution characteristics.}

\section{Experimental Setup}\label{sec:sec4}

This section describes the experimental environment, model configurations, prompt strategy execution, and measurement pipeline used to ensure fair and reproducible evaluation across all prompt strategies and language models \footnote{All experimental code, logs, coverage reports, and derived results are available at \url{https://github.com/Kumari-Pragati/Sustanability_SLMs_Study}.}, with the fixed generation and runtime parameters summarized in Table~\ref{tab:gen_run_params}.

\subsection{Models and Prompt Strategies}

The experiments evaluate three open-source SLMs—\\
\texttt{(i) deepseek-coder-7b-instruct-v1.5}, \texttt{(ii) Meta-Llama-\\-3-8B-Instruct}, and \texttt{(iii) Mistral-7B-Instruct-v0.3} across seven prompt strategies. Each model is executed under identical generation parameters to ensure fairness and isolate the impact of prompt strategy selection. The prompt strategies include Zero-shot, Few-shot, Chain-of-Thought, Least-to-Most, Program-of-Thought, Self-Consistency, and ReAct. For each model–strategy combination, only the prompt formulation and minimal strategy-specific logic are modified, while all other parameters remain fixed.

\subsection{Execution Environment and Tooling}

All experiments are conducted on Google Colab using an NVIDIA A100 GPU with approximately 80\,GB of GPU memory. The experimental environment employs the \texttt{transformers} and \texttt{accelerate} libraries for model loading and inference, \texttt{bitsandbytes} for low-bit quantization, \texttt{tqdm} for batch-level progress tracking, and \texttt{CodeCar-\\-bon} for energy and emissions monitoring. All models are loaded using 4-bit quantization with NF4 configuration to reflect resource-constrained inference settings. Inference is executed using the Hugging Face \texttt{generate} API under a fixed configuration across all runs.

\subsection{Generation Configuration}

Each model invocation uses identical generation parameters. Sampling is enabled with a temperature of 0.2 and nucleus sampling with top-$p$ set to 0.9. The maximum number of generated tokens is capped at 1024 per invocation. Tokenization parallelism is disabled to avoid non-deterministic runtime behavior.
Inference is executed in batches, where each batch consists of 10 functional test generation executions. All executions within a batch share a single CodeCarbon monitoring session to capture cumulative energy, emissions, and runtime statistics.

\subsection{Energy, Emissions, and Token Tracking}

Energy consumption and carbon emissions are tracked using CodeCarbon at the batch level. For each batch execution, CodeCarbon records CPU, GPU, and RAM energy consumption (in kWh), total carbon emissions (in kg CO$_2$ equivalent), and execution duration (in sec). Energy values are later aggregated using summation to compute total energy consumption per model–prompt strategy combination. Token counts are recorded after each model completion by explicitly tracking input tokens, output tokens, and total tokens. These token statistics are logged independently and later merged with energy and emissions data for metric computation.
\begin{table}[t]
\centering
\caption{Generation and Run Parameters}
\label{tab:gen_run_params}
\begin{tabular}{ll}
\hline
\textbf{Parameter} & \textbf{Value} \\
\hline
Temperature & 0.2 \\
Max new tokens & 1024 \\
Batch size & 10 executions per batch \\
Model quantization & 4-bit (bitsandbytes) \\
Top-$p$ & 0.9 \\
Tracking tool & CodeCarbon \\
Token counting & Hugging Face tokenizer \\
\hline
\end{tabular}
\end{table}

\subsection{Run Structure and Data Consolidation}

Each of the three models is evaluated under seven prompt strategies using a separate execution pipeline, yielding a total of 21 \texttt{CodeCarbon} CSV logs. For each model strategy pair, the corresponding CSV file contains 98 batch-level executions, providing sufficient granularity for subsequent summation and normalization of sustainability metrics.

All 21 CSV logs are merged into a unified dataset containing execution time, energy consumption across CPU, GPU, and RAM, carbon emissions, and token statistics. Coverage is computed separately using \texttt{coverage.py} on the generated test suites and normalized to the range $[0,1]$ before being combined with sustainability metrics as described in Section~3.

This consolidated dataset serves as the basis for computing cost-normalized metrics, {\color{black}coverage} quality  efficiency measures, and the sustainability--{\color{black}coverage} quality trade-off score.

\section{Results}
\label{sec:results}\label{sec:sec5}

This section presents an empirical evaluation of the sustainability and efficiency behavior of different prompt strategies applied to small language models for automated test generation. The results are organized to mirror the structure of the framework (Section~\ref{sec:sec3}), enabling direct alignment between measured parameters, derived metrics, and observed outcomes.

We first analyze execution-level behavior, including token usage, runtime, energy consumption, carbon emissions, and normalized test coverage. These primary observations establish the baseline upon which efficiency-oriented and trade-off analyses are later constructed. Subsequent subsections then examine token efficiency, {\color{black}coverage} quality-normalized efficiency, and sustainability--{\color{black}coverage} quality trade-offs in a progressive manner.

\subsection{Execution Time, Energy Consumption, and Coverage Characteristics}
\label{subsec:primary-execution}

This subsection analyzes the execution-level behavior of different prompt strategies based on execution time ($\tau_{hr}$, Eq.~(3)), total energy consumption ($E_{tot}$, Eq.~(2)), and normalized test coverage ($Q$, Eq.~(1)). These metrics are derived from the primary measurements defined in the framework and characterize the baseline sustainability and {\color{black}coverage} test quality of automated test generation prior to token-normalized analysis.

\begin{figure*}[t]
  \centering

  \begin{subfigure}[t]{0.32\textwidth}
    \centering
    \includegraphics[width=\linewidth]{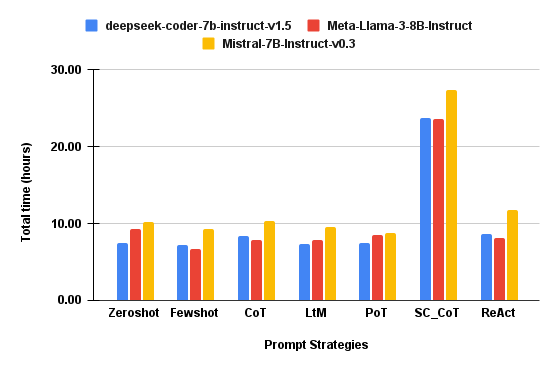}
    \caption{$\tau_{hr}$}
    \label{fig:tau}
  \end{subfigure}
  \hfill
  \begin{subfigure}[t]{0.32\textwidth}
    \centering
    \includegraphics[width=\linewidth]{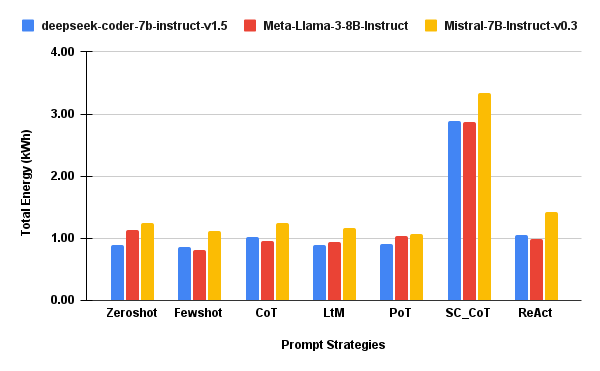}
    \caption{$E_{tot}$}
    \label{fig:energy}
  \end{subfigure}
  \hfill
  \begin{subfigure}[t]{0.32\textwidth}
    \centering
    \includegraphics[width=\linewidth]{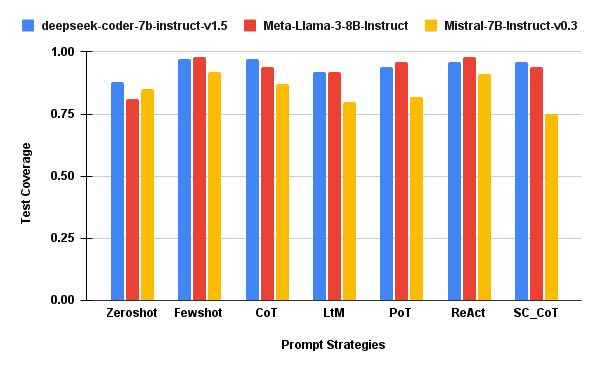}
    \caption{$Q$}
    \label{fig:coverage}
  \end{subfigure}

  \caption{Execution time, energy consumption, and normalized coverage characteristics observed across different prompt strategies.
}
  \label{fig:sustainability-metrics}
\end{figure*}

\textit{Observations.}
As shown in Fig.~\ref{fig:tau}, prompt strategies exhibit substantial variation in execution time across all evaluated models. Lightweight strategies such as \texttt{Zeroshot} and \texttt{Fewshot} consistently achieve the shortest execution times, completing within $6.69$--$10.19\,\mathrm{h}$ across the three SLMs. Moderately structured strategies, including \texttt{LtM}, \texttt{PoT}, and \texttt{ReAct}, incur slightly higher execution times, generally ranging between $7.28$ and $11.71\,\mathrm{h}$. Standard reasoning via \texttt{CoT} remains in a similar band ($7.85$--$10.29\,\mathrm{h}$), whereas the reasoning-intensive \texttt{SC\_CoT} strategy substantially increases runtime for all models, reaching $23.65$--$27.38\,\mathrm{h}$.

Total energy consumption follows a similar pattern, as illustrated in Fig.~\ref{fig:energy}. \texttt{Zeroshot}, \texttt{Fewshot}, \texttt{LtM}, and \texttt{PoT} consistently remain below $1.24\,\mathrm{kWh}$ across models, while \texttt{CoT} is slightly higher but still moderate ($0.95$--$1.25\,\mathrm{kWh}$). \texttt{ReAct} incurs a further increase ($0.99$--$1.42\,\mathrm{kWh}$). In contrast, \texttt{SC\_CoT} produces the highest energy demand for all SLMs, ranging from $2.88$ to $3.34\,\mathrm{kWh}$. These results indicate that increasing reasoning complexity most notably through \texttt{SC\_CoT} incurs a clear energy overhead that is consistent across models.

In contrast, normalized test coverage varies within a comparatively narrower range across prompt strategies, as shown in Fig.~\ref{fig:coverage}. \texttt{Zeroshot} yields the lowest coverage values ($Q=0.81$--$0.88$), while \texttt{Fewshot} achieves the highest overall coverage ($Q=0.92$--$0.98$). \texttt{CoT} and \texttt{ReAct} remain competitive ($Q=0.87$--$0.97$ and $Q=0.91$--$0.98$, respectively). \texttt{LtM} and \texttt{PoT} provide moderate coverage ($Q=0.80$--$0.92$ and $Q=0.82$--$0.96$), while \texttt{SC\_CoT} does not consistently improve coverage despite its high cost, ranging from $Q=0.75$ to $0.96$ depending on the model.

These results indicate that although structured prompt strategies can marginally improve coverage, the gains are not proportional to the substantial increases in execution time and energy demand observed for computationally intensive strategies. This suggests diminishing returns in coverage when increasing reasoning complexity at the execution level.

Overall, these results establish execution time and energy consumption as the primary differentiators among prompt strategies at this stage, while coverage remains relatively robust across strategies. This motivates the subsequent analysis of token-normalized efficiency and {\color{black}coverage} quality-aware sustainability trade-offs.

\subsection{Token-Normalized Efficiency Metrics}
\label{subsec:token-efficiency}

This subsection evaluates execution efficiency by normalizing computational cost with respect to generated output tokens. Using the token-normalized metrics namely token throughput (\textit{TokRate}, Eq.~(4)), time per $1{,}000$ tokens (\textit{SecPer1KTok}, Eq.~(5)), carbon emissions per $1{,}000$ tokens (\textit{CO$_2$Per1KTok}, Eq.~(6)), and energy per $1{,}000$ tokens (\textit{EPer1KTok}, Eq.~(7)), we analyze how different prompt strategies convert computation time, energy, and carbon emissions into usable test artifacts, with observations verified across multiple SLMs.

\begin{figure}[t]
  \centering
  % Line or grouped bar chart: Tokens per hour vs Prompt Strategy per Model
  \includegraphics[width=0.32\textwidth]{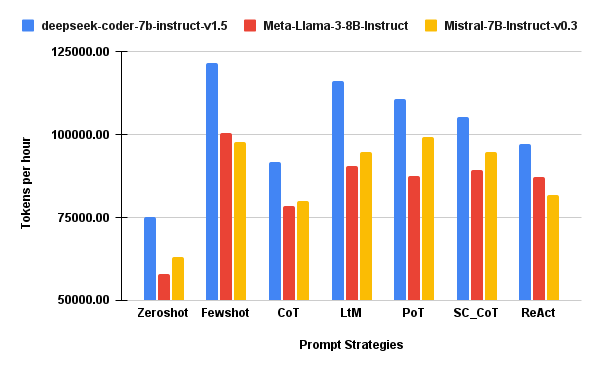}
  \caption{TokRate (TokenThroughput.)}
  \label{fig:token-throughput}
\end{figure}

\textit{Observations.}
As shown in Fig.~\ref{fig:token-throughput}, token throughput (\textit{TokRate}) varies substantially across prompt strategies, with consistent trends observed across all evaluated models. \texttt{Zeroshot} prompting consistently yields lower throughput, while lightweight and moderately structured strategies such as \texttt{Fewshot}, \texttt{LtM}, and \texttt{PoT} achieve higher token generation rates. For example, under \texttt{deepseek-coder-7b}, throughput increases from $75{,}299.25$ tokens/hour for \texttt{Zeroshot} to $121{,}733.04$, $116{,}104.42$, and $110{,}801.95$ tokens/hour for \texttt{Fewshot}, \texttt{LtM}, and \texttt{PoT}, respectively, while reasoning-intensive strategies such as \texttt{SC\_CoT} and \texttt{ReAct} reduce throughput. The same prompt-level ordering is reflected for \texttt{Meta-Llama-3-8B-Instruct} and \texttt{Mistral-7B-Instruct-v0.3}. These trends indicate that deeper reasoning chains reduce effective token generation speed despite increased computational effort.

\begin{figure*}[t]
  \centering

  \begin{subfigure}[t]{0.32\textwidth}
    \centering
    \includegraphics[width=\linewidth]{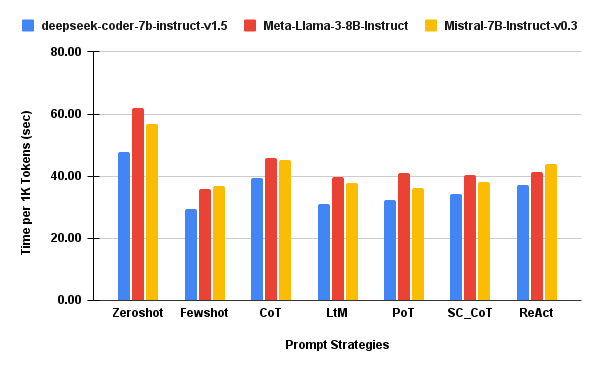}
    \caption{SecPer1KTok}
    \label{fig:sec-per-1k}
  \end{subfigure}
  \hfill
  \begin{subfigure}[t]{0.32\textwidth}
    \centering
    \includegraphics[width=\linewidth]{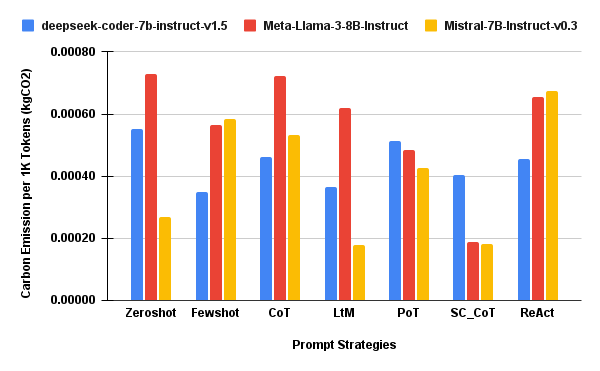}
    \caption{$CO_2$Per1KTok}
    \label{fig:co2-per-1k}
  \end{subfigure}
  \hfill
  \begin{subfigure}[t]{0.32\textwidth}
    \centering
    \includegraphics[width=\linewidth]{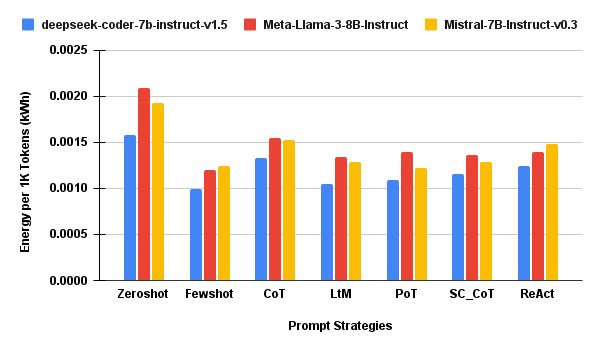}
    \caption{EPer1KTok}
    \label{fig:energy-per-1k}
  \end{subfigure}

  \caption{Normalized sustainability metrics per 1K generated tokens:
  (a) execution time in seconds, (b)  carbon emissions, and (c) energy consumption.}
  \label{fig:per-1k-metrics}
\end{figure*}

The time cost per $1{,}000$ tokens (\textit{SecPer1KTok}, Fig.~\ref{fig:sec-per-1k}) further reinforces prompt-driven differences in execution efficiency. Across all models, \texttt{Fewshot} and \texttt{LtM} consistently exhibit the lowest latency, while \texttt{PoT} and \texttt{CoT} incur moderate time costs. In contrast, \texttt{SC\_CoT}, \texttt{ReAct}, and especially \texttt{Zeroshot} introduce higher latency. For instance, under \texttt{deepseek-coder-7b}, latency decreases from $47.81\,\mathrm{s}$ for \texttt{Zeroshot} to $29.57\,\mathrm{s}$ for \texttt{Fewshot}, while \texttt{SC\_CoT} increases latency to $34.17\,\mathrm{s}$. Similar prompt-level patterns are observed for \texttt{Meta-Llama-3-8B-Instruct} and \texttt{Mistral-7B-Instruct-v0.3}, 

\noindent confirming that reasoning depth and prompt structure directly influence token-level latency independent of the underlying model.

Carbon emissions per $1{,}000$ tokens (\textit{CO$_2$Per1KTok}, Fig.~\ref{fig:co2-per-1k}) show clear differences across prompt strategies, with consistent relative behavior across models. Lightweight strategies such as \texttt{Fewshot} and \texttt{LtM} consistently incur lower emissions, whereas \texttt{ReAct}, \texttt{SC\_CoT}, and \texttt{Zeroshot} exhibit higher carbon costs. For example, under \texttt{deepseek-coder-7b}, emissions decrease from $0.00055\,\mathrm{kg}$ for \texttt{Zero-\\-shot} to $0.00035\,\mathrm{kg}$ for \texttt{Fewshot}, while \texttt{SC\_CoT} increases emissions to $0.00040\,\mathrm{kg}$. Comparable prompt-level differences are observed for \texttt{Meta-Llama-3-8B-Instruct} and \texttt{Mistral-7B-Instruct-v0.3}, 

\noindent demonstrating that prompt strategy selection materially affects token-level carbon cost across models.

Energy per $1{,}000$ tokens (\textit{EPer1KTok}, Fig.~\ref{fig:energy-per-1k}) follows a similar prompt-driven pattern. Across all evaluated models, \texttt{Fewshot} and \texttt{LtM} maintain the lowest energy consumption, while \texttt{PoT} and \texttt{CoT} incur moderate costs. In contrast, \texttt{SC\_CoT} and \texttt{ReAct} consistently occupy the higher end of the energy spectrum. For instance, under \texttt{Meta-Llama-3-8B-Instruct}, energy usage decreases from $0.0021\,\mathrm{kWh}$ for \texttt{Zeroshot} to approximately $0.0012$–$0.0013\,\mathrm{kWh}$ for lightweight strategies, while \texttt{SC\_CoT} increases energy consumption to $0.0014\,\mathrm{kWh}$. The same ordering of prompt strategies is observed for \texttt{deepseek-coder-7b} and \texttt{Mistral-7B-Instruct-v0.3}.

Overall, token-normalized efficiency metrics consistently show that structured but lightweight prompt strategies deliver higher output efficiency with lower time, energy, and carbon cost per unit of generated content across all evaluated models, as reflected in Figs.~\ref{fig:token-throughput}--\ref{fig:per-1k-metrics}. In contrast, reasoning-intensive strategies such as \texttt{SC\_CoT} and \texttt{ReAct}, as well as unstructured \texttt{Zeroshot} prompting, incur higher sustainability costs without proportional gains in token-level efficiency. These consolidated observations provide the analytical grounding for the subsequent analysis of how such efficiency differences translate into {\color{black}coverage} testing quality and sustainability--{\color{black}coverage} quality trade-offs.

\begin{figure*}[t]
  \centering

  \begin{subfigure}[t]{0.32\textwidth}
    \centering
    \includegraphics[width=\linewidth]{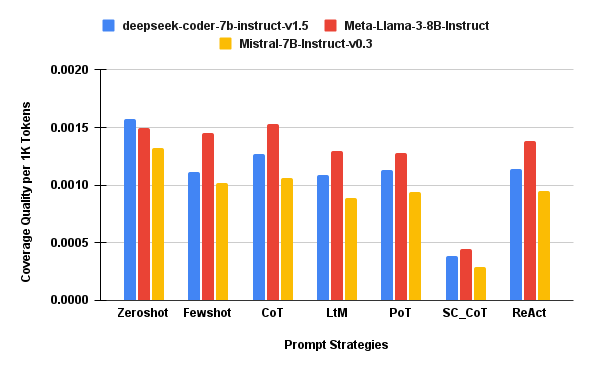}
    \caption{QPer1KTok}
    \label{fig:q-per-1k}
  \end{subfigure}
  \hfill
  \begin{subfigure}[t]{0.32\textwidth}
    \centering
    \includegraphics[width=\linewidth]{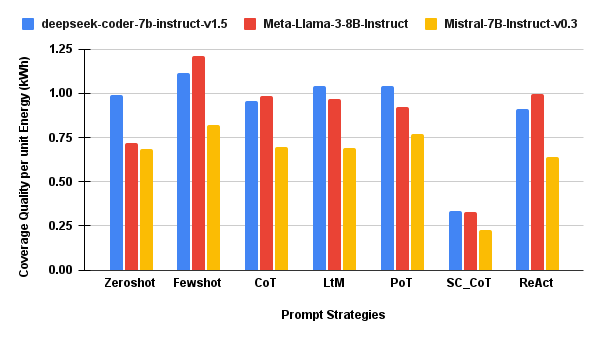}
    \caption{QPerkWh}
    \label{fig:q-per-kwh}
  \end{subfigure}
  \hfill
  \begin{subfigure}[t]{0.32\textwidth}
    \centering
    \includegraphics[width=\linewidth]{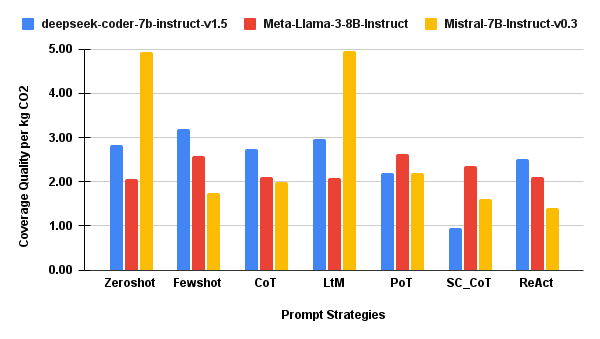}
    \caption{QPer$CO_2$}
    \label{fig:q-per-co2}
  \end{subfigure}

  \caption{{\color{black}coverage} quality-efficiency metrics across prompt strategies:
  (a) normalized coverage per 1K tokens, (b) coverage per kWh, and (c) coverage per $CO_2$ emissions.}
  \label{fig:quality-efficiency}
\end{figure*}

\subsection{{\color{black}Coverage} Quality Efficiency Metrics}
\label{subsec:quality-efficiency}

This subsection analyzes how effectively different prompt strategies translate computational and environmental resources into {\color{black}coverage} testing quality. Quality is measured using normalized test coverage ($Q$) and evaluated through three quality-efficiency metrics: {\color{black}coverage} quality per $1{,}000$ tokens (\textit{QPer1KTok}, Eq.~(8)), {\color{black}coverage} quality per kilowatt-hour (\textit{QPerkWh}, Eq.~(9)), and {\color{black}coverage} quality per kilogram of $\mathrm{CO}_2$ (\textit{QPerCO$_2$}, Eq.~(10)). Together, these metrics assess whether increased computational cost yields proportionate improvements in testing effectiveness.

{\color{black}Coverage} Quality per 1,000 tokens (\textit{QPer1KTok,}  
Fig~\ref{fig:q-per-1k}) reports coverage efficiency normalized by token usage. Across all evaluated models, lightweight and moderately structured strategies achieve higher quality per $1{,}000$ tokens than reasoning-intensive approaches. For \texttt{Meta-Llama-3-8B-Instruct}, \texttt{Zeroshot}, \texttt{Fewshot}, and \texttt{CoT} attain comparable peak values around $0.0015$, whereas \texttt{SC\_CoT} drops sharply to approximately $0.0004$. A similar reduction under \texttt{SC\_CoT} is observed for \texttt{deepseek-coder-7b} ($0.0004$) and \texttt{Mistral-7B-Instruct-v0.3} ($0.0003$), indicating that increased token volume from self-consistency does not translate into proportional coverage gains.

{\color{black}Coverage} Quality per kilowatt-hour (\textit{QPerkWh, Fig.~\ref{fig:q-per-kwh}}) reports energy-normalized coverage efficiency. Energy-normalized {\color{black}coverage} quality further highlights sustainability differences among prompt strategies. \texttt{Fewshot} achieves the highest {\color{black}coverage} quality per unit energy for \texttt{deepseek-coder-7b} ($1.12\,Q/\mathrm{kWh}$) and \texttt{Meta-\\-Llama-3-8B-Instruct} ($1.21\,Q/\mathrm{kWh}$), while \texttt{LtM} and \texttt{PoT} also remain energy efficient across models. In contrast, \texttt{SC\_CoT} consistently yields the lowest energy-normalized {\color{black}coverage} quality, with values around $0.33\,Q/\mathrm{kWh}$ for \texttt{deepseek-coder-7b} and \texttt{Meta-\\-Llama-3-8B-Instruct} and $0.22\,Q/\mathrm{kWh}$ for \texttt{Mistral-7B-Inst-\\-ruct-v0.3}, reflecting poor energy efficiency relative to coverage gains.

{\color{black}Coverage} Quality per kilogram of $\mathrm{CO}_2$ (\textit{QPerCO$_2$, Fig.~\ref{fig:q-per-co2}}) reports carbon-normalized coverage efficiency. Carbon-normalized {\color{black}coverage} quality shows strong separation between prompt strategies. Lightweight strategies such as \texttt{Fewshot}, \texttt{LtM}, and \texttt{Zero-\\-shot} achieve high coverage per kilogram of emitted $\mathrm{CO}_2$, exceeding $3.0\,Q/\mathrm{kg}$ for \texttt{deepseek-coder-7b} and reaching values close to $5.0\,Q/\mathrm{kg}$ for \texttt{Mistral-7B-Instruct-v0.3}. While \texttt{SC\_CoT} yields the lowest carbon-normalized {\color{black}coverage} quality for \texttt{deepseek-\\-coder-7b} ($0.95\,Q/\mathrm{kg}$), it remains substantially less efficient than lightweight strategies for \texttt{Meta-Llama-3-8B-Instruct} and \texttt{Mis-\\-tral-7B-Instruct-v0.3} as well, indicating that its increased emissions are not offset by commensurate coverage improvements.

Overall, {\color{black}coverage} quality efficiency metrics consistently demonstrate that structured yet lightweight prompt strategies provide a superior sustainability--{\color{black}coverage} quality trade-off. In contrast, reasoning-intensive strategies such as \texttt{SC\_CoT} substantially increase token usage, energy consumption, and carbon emissions without delivering proportional gains in {\color{black}coverage} testing quality, underscoring the importance of prompt design for sustainable automated software testing.

\subsection{Sustainability--{\color{black}Coverage} Quality Trade-off Analysis}
\label{subsec:tradeoff}

\begin{figure*}[t]
  \centering

  \begin{subfigure}[t]{0.32\textwidth}
    \centering
    \includegraphics[width=\linewidth]{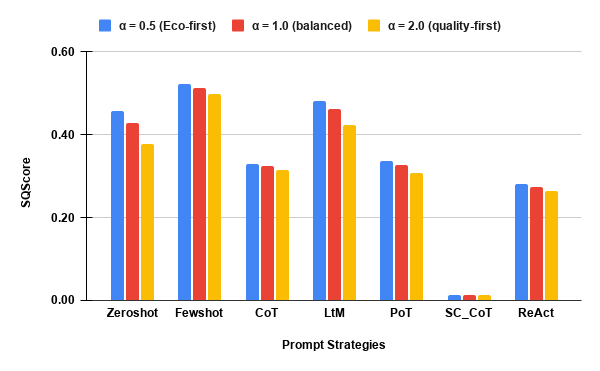}
    \caption{deepseek-coder-7b-instruct-v1.5}
    \label{fig:sqscore-deepseek}
  \end{subfigure}
  \hfill
  \begin{subfigure}[t]{0.32\textwidth}
    \centering
    \includegraphics[width=\linewidth]{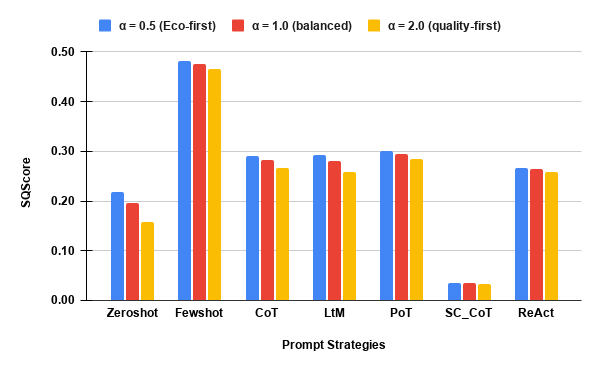}
    \caption{Meta-Llama-3-8B-Instruct}
    \label{fig:sqscore-meta}
  \end{subfigure}
  \hfill
  \begin{subfigure}[t]{0.32\textwidth}
    \centering
    \includegraphics[width=\linewidth]{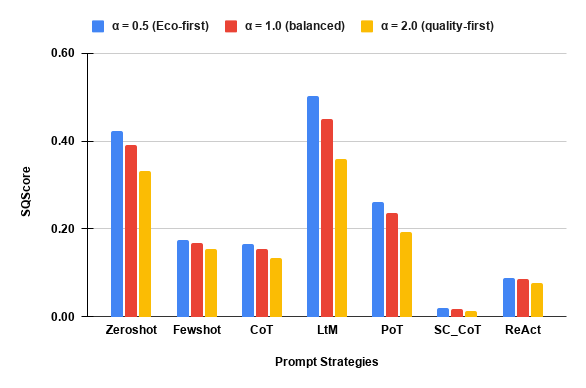}
    \caption{Mistral-7B-Instruct-v0.3}
    \label{fig:sqscore-mistral}
  \end{subfigure}

  \caption{SQScore comparison across small language models under different prompt strategies: (a) DeepSeek-Coder-7B, (b)
Meta-Llama-3-8B, and (c) Mistral-7B.}
  \label{fig:sqscore-comparison}
\end{figure*}

This subsection evaluates the sustainability--{\color{black}coverage} quality trade-off using the composite SQScore defined in Eq.~(11). Results are reported for three prioritization regimes—eco-first ($\alpha = 0.5$), balanced ($\alpha = 1.0$), and quality-first ($\alpha = 2.0$)—to analyze how different prompt strategies behave under varying sustainability preferences.

\textit{Observations.}
The results for \texttt{deepseek-coder-7b-instruct-\\-v1.5} (Fig.~\ref{fig:sqscore-deepseek}) show that \texttt{Fewshot} consistently achieves the highest \textit{SQScore} across all settings ($0.52$ at $\alpha=0.5$, $0.51$ at $\alpha=1.0$, and $0.50$ at $\alpha=2.0$). \texttt{Zeroshot} and \texttt{LtM} form the next tier, with \texttt{Zeroshot} decreasing from $0.46$ to $0.38$ as $\alpha$ increases and \texttt{LtM} decreasing from $0.48$ to $0.42$. In contrast, \texttt{SC\_CoT} remains the lowest ($0.01$ for all $\alpha$), indicating that its increased execution overhead dominates any marginal quality benefit under the composite score.

The results for \texttt{Meta-Llama-3-8B-Instruct} (Fig.~\ref{fig:sqscore-meta}) show that \texttt{Fewshot} yields the strongest trade-off, with a slight decrease from $0.48$ at $\alpha = 0.5$ to $0.47$ at $\alpha = 1.0$ and $0.47$ at $\alpha = 2.0$. \texttt{PoT} is consistently the second-best option (from $0.30$ at $\alpha = 0.5$ to $0.28$ at $\alpha = 2.0$), while \texttt{CoT} and \texttt{LtM} cluster in a similar range (approximately $0.29$ down to $0.26$). As in the other models, \texttt{SC\_CoT} produces the lowest \textit{SQScore} ($0.04$, $0.03$, and $0.03$), reflecting an unfavorable sustainability--{\color{black}coverage} quality balance.

The results for \texttt{Mistral-7B-Instruct-v0.3} (Fig.~\ref{fig:sqscore-mistral}) show that \texttt{LtM} achieves the highest \textit{SQScore} under all settings, decreasing from $0.50$ (eco-first) to $0.36$ (quality-first). \texttt{Zeroshot} ranks second and shows a clearer decline with increasing $\alpha$ ($0.42 \rightarrow 0.33$). Other strategies remain comparatively lower (e.g., \texttt{PoT}: $0.26 \rightarrow 0.19$; \texttt{Fewshot}: $0.17 \rightarrow 0.15$), while \texttt{SC\_CoT} again yields the minimum scores ($0.02$, $0.02$, and $0.01$).

Overall, the \textit{SQScore} analysis (Eq.~(11)) shows that the best sustainability {\color{black}coverage} quality trade-off is typically achieved by lightweight yet structured prompting, with \texttt{Fewshot} dominating for \texttt{deepseek-cod- er-7b-instruct-v1.5} and \texttt{Meta-Llama-3-8B-\\-Instruct}, and \texttt{LtM} dominating for \texttt{Mistral-7B-Instruct-v0.3}. Across all three models and prioritization settings, \texttt{SC\_CoT} consistently ranks last, indicating that its added computational and environmental cost is not compensated by commensurate quality gains within the composite trade-off formulation.

\section{Discussion}
\label{sec:sec6}

The experimental results show that prompt strategies can induce markedly different computational and environmental behaviors for the same testing task, even when the model and dataset remain fixed. This indicates that prompt formulation itself acts as a first-order factor in shaping execution dynamics and sustainability characteristics. From an interpretive perspective, these differences arise from how
reasoning structure influences model behavior. Lightweight and
moderately structured prompts tend to encourage more direct solution
paths, resulting in lower execution overhead. However, deeper
reasoning formulations introduce additional generation and
verification steps that amplify runtime, energy usage, and emissions.

The results summarized in Table~\ref{tab:overall-summary} indicate that these increased costs do not consistently translate into proportional improvements in {\color{black}coverage} testing quality.
 Strategies such as \texttt{PoT} and \texttt{ReAct} illustrate this trade-off, where structured reasoning can enhance interpretability and control but at the expense of reduced efficiency under sustainability-aware evaluation. Conversely, \texttt{SC\_CoT} demonstrates that increasing reasoning depth beyond a certain point leads to diminishing returns, with environmental and execution costs dominating overall outcomes. The consistency of these patterns across models suggests that
sustainability-sensitive behavior is driven more by prompt-induced
reasoning characteristics than by architectural differences alone.
This indeed implies that prompt engineering has the potential to
emerge as a practical mechanism for controlling the environmental
footprint of automated software testing workflows.

 \begin{table}[t]
\centering
\caption{Overall sustainability and {\color{black}coverage} quality summary across prompt strategies and models.}
\label{tab:overall-summary}
\scriptsize
\begin{tabular}{l l c c c c}
\toprule
\textbf{Model} & \textbf{Prompt} & $Q$ & $E_{\text{1K}}$ & $C_{\text{1K}}$ & $\mathrm{SQScore}_{\alpha=1.0}$ \\
\midrule

\texttt{Meta-Llama-3-8B-Instruct} 
& \texttt{Zeroshot} & $0.81$ & $0.0021$ & $0.00073$ & $0.20$ \\
& \texttt{Fewshot} & $0.98$ & $0.0012$ & $0.00057$ & $0.48$ \\
& \texttt{CoT}     & $0.94$ & $0.0015$ & $0.00072$ & $0.28$ \\
& \texttt{LtM}     & $0.92$ & $0.0013$ & $0.00062$ & $0.28$ \\
& \texttt{PoT}     & $0.96$ & $0.0014$ & $0.00049$ & $0.30$ \\
& \texttt{SC\_CoT} & $0.94$ & $0.0014$ & $0.00019$ & $0.03$ \\
& \texttt{ReAct}   & $0.98$ & $0.0014$ & $0.00065$ & $0.26$ \\
\midrule

\texttt{DeepSeek-Coder-7B-Instruct} 
& \texttt{Zeroshot} & $0.88$ & $0.0016$ & $0.00055$ & $0.43$ \\
& \texttt{Fewshot} & $0.97$ & $0.0010$ & $0.00035$ & $0.51$ \\
& \texttt{CoT}     & $0.97$ & $0.0013$ & $0.00046$ & $0.32$ \\
& \texttt{LtM}     & $0.92$ & $0.0010$ & $0.00037$ & $0.46$ \\
& \texttt{PoT}     & $0.94$ & $0.0011$ & $0.00051$ & $0.33$ \\
& \texttt{SC\_CoT} & $0.98$ & $0.0012$ & $0.00040$ & $0.01$ \\
& \texttt{ReAct}   & $0.98$ & $0.0012$ & $0.00045$ & $0.27$ \\
\midrule

\texttt{Mistral-7B-Instruct} 
& \texttt{Zeroshot} & $0.85$ & $0.0019$ & $0.00027$ & $0.39$ \\
& \texttt{Fewshot} & $0.92$ & $0.0012$ & $0.00058$ & $0.17$ \\
& \texttt{CoT}     & $0.87$ & $0.0015$ & $0.00053$ & $0.15$ \\
& \texttt{LtM}     & $0.80$ & $0.0013$ & $0.00018$ & $0.45$ \\
& \texttt{PoT}     & $0.82$ & $0.0012$ & $0.00043$ & $0.24$ \\
& \texttt{SC\_CoT} & $0.75$ & $0.0013$ & $0.00018$ & $0.02$ \\
& \texttt{ReAct}   & $0.91$ & $0.0015$ & $0.00068$ & $0.08$ \\
\bottomrule
\end{tabular}
\end{table}

\section{Threats to Validity}
\label{sec:sec7}

It is important to highlight some potential threats to the validity of the results presented in the previous sections.

\textit{Internal Validity:}
Although the experimental pipeline was kept constant across all prompt strategies and models, uncontrolled system-level factors such as background processes, model loading variability, caching effects, and runtime scheduling may have influenced execution time, energy consumption, and token measurements. All experiments were conducted with the system continuously connected to a power source to mitigate battery-related fluctuations; however, minor variations in power draw or thermal behavior may still introduce small measurement biases. {\color{black}Additionally, stochasticity in model generation may introduce variability across runs despite controlled configurations.}

\textit{External Validity:}
The generalizability of the results is limited by the use of a single benchmark dataset (MBPP), a fixed hardware configuration, and assumed electricity carbon intensity values. While MBPP supports controlled evaluation of automated test generation, sustainability and efficiency trends may differ for {\color{black}larger software systems, alternative datasets, programming languages, or real-world industrial environments}. Results may also vary across hardware platforms, geographic regions, or execution settings{\color{black}, particularly where energy infrastructure and carbon intensity differ.}

\textit{Construct Validity:}
Energy consumption and carbon emissions were estimated using CodeCarbon, which relies on standardized carbon intensity factors rather than direct power-meter readings. Consequently, reported emission values represent approximations. Token usage was aggregated from input and output tokens during batch execution and serves as a proxy for inference cost, but does not capture all low-level hardware or runtime optimizations. Test quality was measured using code coverage via \texttt{coverage.py}, which {\color{black}reflects structural test adequacy but does not fully capture semantic correctness, fault-detection effectiveness, or test oracle strength}. {\color{black}Therefore, quality-related conclusions in this study are limited to coverage evaluation.}

\textit{Statistical Conclusion Validity:}
Experiments were conducted over a finite set of MBPP tasks and aggregated at the batch level for each model–prompt combination. As a result, stochastic variation in model inference and system execution may affect individual measurements. Minor differences in sustainability and efficiency metrics may not reach statistical significance{\color{black}, as formal statistical tests and confidence intervals were not explicitly evaluated}, and observed correlations should therefore be interpreted with caution.

\section{Conclusion and Future Work}
\label{sec:sec8}

This work presented a systematic sustainability--{\color{black}coverage} quality evaluation of prompt engineering strategies for automated unit test generation using the MBPP benchmark. The proposed framework integrates execution behavior, environmental impact, and {\color{black}coverage} testing quality into a unified assessment, enabling prompt strategies to be compared beyond coverage alone.

The experimental results reveal strong prompt-dependent sustainability patterns. Lightweight strategies such as \texttt{Zero-shot} and \texttt{Few-shot} consistently exhibit lower energy consumption and carbon emissions, while reasoning-intensive approaches, particularly \texttt{Self-Consistency} and \texttt{Chain-of-Thought}, achieve marginally higher coverage at substantially increased computational and environmental cost. These findings demonstrate that prompt design alone can significantly influence sustainability outcomes, largely independent of the underlying model, and that no single prompt strategy optimizes all efficiency and {\color{black}coverage} quality dimensions simultaneously.

Future work will extend this evaluation to broader software benchmarks, explore additional quality indicators beyond code coverage, and investigate adaptive prompt selection approaches that dynamically balance testing effectiveness with sustainability constraints in practical automated testing pipelines.

%%
%% The next two lines define the bibliography style to be used, and
%% the bibliography file.
\bibliographystyle{ACM-Reference-Format}
\bibliography{sample-base}

%%% -*-BibTeX-*-
%%% Do NOT edit. File created by BibTeX with style
%%% ACM-Reference-Format-Journals [18-Jan-2012].

\begin{thebibliography}{25}

%%% ====================================================================
%%% NOTE TO THE USER: you can override these defaults by providing
%%% customized versions of any of these macros before the \bibliography
%%% command.  Each of them MUST provide its own final punctuation,
%%% except for \shownote{} and \showURL{}.  The latter two
%%% do not use final punctuation, in order to avoid confusing it with
%%% the Web address.
%%%
%%% To suppress output of a particular field, define its macro to expand
%%% to an empty string, or better, \unskip, like this:
%%%
%%% \newcommand{\showURL}[1]{\unskip}   % LaTeX syntax
%%%
%%% \def \showURL #1{\unskip}           % plain TeX syntax
%%%
%%% ====================================================================

\ifx \showCODEN    \undefined \def \showCODEN     #1{\unskip}     \fi
\ifx \showISBNx    \undefined \def \showISBNx     #1{\unskip}     \fi
\ifx \showISBNxiii \undefined \def \showISBNxiii  #1{\unskip}     \fi
\ifx \showISSN     \undefined \def \showISSN      #1{\unskip}     \fi
\ifx \showLCCN     \undefined \def \showLCCN      #1{\unskip}     \fi
\ifx \shownote     \undefined \def \shownote      #1{#1}          \fi
\ifx \showarticletitle \undefined \def \showarticletitle #1{#1}   \fi
\ifx \showURL      \undefined \def \showURL       {\relax}        \fi
% The following commands are used for tagged output and should be
% invisible to TeX
\providecommand\bibfield[2]{#2}
\providecommand\bibinfo[2]{#2}
\providecommand\natexlab[1]{#1}
\providecommand\showeprint[2][]{arXiv:#2}

\bibitem[Cheng et~al\mbox{.}(2025)]%
        {cheng2025revisiting}
\bibfield{author}{\bibinfo{person}{Xiang Cheng}, \bibinfo{person}{Chengyan Pan}, \bibinfo{person}{Minjun Zhao}, \bibinfo{person}{Deyang Li}, \bibinfo{person}{Fangchao Liu}, \bibinfo{person}{Xinyu Zhang}, \bibinfo{person}{Xiao Zhang}, {and} \bibinfo{person}{Yong Liu}.} \bibinfo{year}{2025}\natexlab{}.
\newblock \showarticletitle{Revisiting Chain-of-Thought Prompting: Zero-shot Can Be Stronger than Few-shot}.
\newblock \bibinfo{journal}{\emph{arXiv preprint}}  \bibinfo{volume}{arXiv:2506.14641} (\bibinfo{year}{2025}).
\newblock
\urldef\tempurl%
\url{https://arxiv.org/abs/2506.14641}
\showURL{%
\tempurl}


\bibitem[Chowdhury and Hindle(2016)]%
        {greenoracle}
\bibfield{author}{\bibinfo{person}{Shaiful~Alam Chowdhury} {and} \bibinfo{person}{Abram Hindle}.} \bibinfo{year}{2016}\natexlab{}.
\newblock \showarticletitle{GreenOracle: Estimating Software Energy Consumption with Energy Measurement Corpora}. In \bibinfo{booktitle}{\emph{Proceedings of the 13th International Conference on Mining Software Repositories (MSR)}}. \bibinfo{pages}{49--60}.
\newblock
\href{https://doi.org/10.1145/2901739.2901763}{doi:\nolinkurl{10.1145/2901739.2901763}}


\bibitem[De~Martino et~al\mbox{.}(2026)]%
        {green_prompt_engineering}
\bibfield{author}{\bibinfo{person}{Vincenzo De~Martino}, \bibinfo{person}{Mohammad~Amin Zadenoori}, \bibinfo{person}{Xavier Franch}, {and} \bibinfo{person}{Alessio Ferrari}.} \bibinfo{year}{2026}\natexlab{}.
\newblock \showarticletitle{Green Prompt Engineering: Investigating the Energy Impact of Prompt Design in Software Engineering}. In \bibinfo{booktitle}{\emph{Proceedings of the ACM Conference on Software Engineering}}.
\newblock
\newblock
\shownote{Pre-print}.


\bibitem[Della~Porta et~al\mbox{.}(2025)]%
        {dellaporta2025promptpatterns}
\bibfield{author}{\bibinfo{person}{Antonio Della~Porta}, \bibinfo{person}{Stefano Lambiase}, {and} \bibinfo{person}{Fabio Palomba}.} \bibinfo{year}{2025}\natexlab{}.
\newblock \showarticletitle{Do Prompt Patterns Affect Code Quality? A First Empirical Assessment of ChatGPT-Generated Code}. In \bibinfo{booktitle}{\emph{Proceedings of the 29th International Conference on Evaluation and Assessment in Software Engineering}}. \bibinfo{publisher}{ACM}, \bibinfo{address}{Istanbul, Turkiye}.
\newblock
\href{https://doi.org/10.1145/3756681.3756938}{doi:\nolinkurl{10.1145/3756681.3756938}}


\bibitem[Durelli et~al\mbox{.}(2025)]%
        {durelli2025_energy_footprint_slm_tests}
\bibfield{author}{\bibinfo{person}{Rafael Durelli}, \bibinfo{person}{Andre Endo}, {and} \bibinfo{person}{Vinicius Durelli}.} \bibinfo{year}{2025}\natexlab{}.
\newblock \showarticletitle{On the Energy Footprint of Using a Small Language Model for Unit Test Generation}.
\newblock \bibinfo{journal}{\emph{Technical Report}} (\bibinfo{year}{2025}).
\newblock
\urldef\tempurl%
\url{https://zenodo.org/records/15809327}
\showURL{%
\tempurl}
\newblock
\shownote{Available on Zenodo}.


\bibitem[Husom et~al\mbox{.}(2024)]%
        {price_of_prompting}
\bibfield{author}{\bibinfo{person}{Erik~Johannes Husom}, \bibinfo{person}{Arda Goknil}, \bibinfo{person}{Lwin~Khin Shar}, {and} \bibinfo{person}{Sagar Sen}.} \bibinfo{year}{2024}\natexlab{}.
\newblock \showarticletitle{The Price of Prompting: Profiling Energy Use in Large Language Models Inference}.
\newblock \bibinfo{journal}{\emph{arXiv preprint arXiv:2407.16893}} (\bibinfo{year}{2024}).
\newblock
\urldef\tempurl%
\url{https://arxiv.org/abs/2407.16893}
\showURL{%
\tempurl}
\newblock
\shownote{Under review}.


\bibitem[Jiang et~al\mbox{.}(2024)]%
        {llm_lifecycle_energy}
\bibfield{author}{\bibinfo{person}{Peng Jiang}, \bibinfo{person}{Christian Sonne}, \bibinfo{person}{Wangliang Li}, \bibinfo{person}{Fengqi You}, {and} \bibinfo{person}{Siming You}.} \bibinfo{year}{2024}\natexlab{}.
\newblock \showarticletitle{Preventing the Immense Increase in the Life-Cycle Energy and Carbon Footprints of LLM-Powered Intelligent Chatbots}.
\newblock \bibinfo{journal}{\emph{Engineering}}  \bibinfo{volume}{40} (\bibinfo{year}{2024}), \bibinfo{pages}{202--210}.
\newblock
\href{https://doi.org/10.1016/j.eng.2024.04.002}{doi:\nolinkurl{10.1016/j.eng.2024.04.002}}


\bibitem[Lemieux et~al\mbox{.}(2023)]%
        {lemieux2023codamosa}
\bibfield{author}{\bibinfo{person}{Caroline Lemieux}, \bibinfo{person}{Jeevana~Priya Inala}, \bibinfo{person}{Shuvendu~K. Lahiri}, {and} \bibinfo{person}{Saurabh Sen}.} \bibinfo{year}{2023}\natexlab{}.
\newblock \showarticletitle{CodaMOSA: Escaping Coverage Plateaus in Test Generation with Pre-trained Large Language Models}. In \bibinfo{booktitle}{\emph{Proceedings of the 45th IEEE/ACM International Conference on Software Engineering (ICSE)}}.
\newblock
\href{https://doi.org/10.1109/ICSE48619.2023.00085}{doi:\nolinkurl{10.1109/ICSE48619.2023.00085}}


\bibitem[Lu et~al\mbox{.}(2021)]%
        {lu2021codexglue}
\bibfield{author}{\bibinfo{person}{Shuai Lu}, \bibinfo{person}{Daya Guo}, \bibinfo{person}{Shuo Ren}, \bibinfo{person}{Junjie Huang}, \bibinfo{person}{Alexey Svyatkovskiy}, \bibinfo{person}{Ambrosio Blanco}, \bibinfo{person}{Colin Clement}, \bibinfo{person}{Dawn Drain}, \bibinfo{person}{Daxin Jiang}, \bibinfo{person}{Duyu Tang}, \bibinfo{person}{Ge Li}, \bibinfo{person}{Lidong Zhou}, \bibinfo{person}{Linjun Shou}, \bibinfo{person}{Long Zhou}, \bibinfo{person}{Michele Tufano}, \bibinfo{person}{Ming Gong}, \bibinfo{person}{Ming Zhou}, \bibinfo{person}{Nan Duan}, \bibinfo{person}{Neel Sundaresan}, \bibinfo{person}{Shao~Kun Deng}, \bibinfo{person}{Shengyu Fu}, {and} \bibinfo{person}{Shujie Liu}.} \bibinfo{year}{2021}\natexlab{}.
\newblock \showarticletitle{CodeXGLUE: A Machine Learning Benchmark Dataset for Code Understanding and Generation}.
\newblock \bibinfo{journal}{\emph{arXiv preprint}}  \bibinfo{volume}{arXiv:2102.04664} (\bibinfo{year}{2021}).
\newblock
\urldef\tempurl%
\url{https://arxiv.org/abs/2102.04664}
\showURL{%
\tempurl}


\bibitem[Pan et~al\mbox{.}(2025)]%
        {hidden_cost_readability}
\bibfield{author}{\bibinfo{person}{Dangfeng Pan}, \bibinfo{person}{Zhensu Sun}, \bibinfo{person}{Cenyuan Zhang}, \bibinfo{person}{David Lo}, {and} \bibinfo{person}{Xiaoning Du}.} \bibinfo{year}{2025}\natexlab{}.
\newblock \showarticletitle{The Hidden Cost of Readability: How Code Formatting Silently Consumes Your LLM Budget}.
\newblock \bibinfo{journal}{\emph{arXiv preprint arXiv:2508.13666}} (\bibinfo{year}{2025}).
\newblock


\bibitem[Pan and Zhang(2025)]%
        {pan2025modularization}
\bibfield{author}{\bibinfo{person}{Ruwei Pan} {and} \bibinfo{person}{Hongyu Zhang}.} \bibinfo{year}{2025}\natexlab{}.
\newblock \showarticletitle{Modularization is Better: Effective Code Generation with Modular Prompting}.
\newblock \bibinfo{journal}{\emph{arXiv preprint}}  \bibinfo{volume}{arXiv:2503.12483} (\bibinfo{year}{2025}).
\newblock
\urldef\tempurl%
\url{https://arxiv.org/abs/2503.12483}
\showURL{%
\tempurl}


\bibitem[Payoungkhamdee et~al\mbox{.}(2025)]%
        {payoungkhamdee2025potmultilingual}
\bibfield{author}{\bibinfo{person}{Patomporn Payoungkhamdee}, \bibinfo{person}{Pume Tuchinda}, \bibinfo{person}{Jinheon Baek}, \bibinfo{person}{Samuel Cahyawijaya}, \bibinfo{person}{Can Udomcharoenchaikit}, \bibinfo{person}{Potsawee Manakul}, \bibinfo{person}{Peerat Limkonchotiwat}, \bibinfo{person}{Ekapol Chuangsuwanich}, {and} \bibinfo{person}{Sarana Nutanong}.} \bibinfo{year}{2025}\natexlab{}.
\newblock \showarticletitle{Towards Better Understanding of Program-of-Thought Reasoning in Cross-Lingual and Multilingual Environments}.
\newblock \bibinfo{journal}{\emph{arXiv preprint}}  \bibinfo{volume}{arXiv:2502.17956} (\bibinfo{year}{2025}).
\newblock
\urldef\tempurl%
\url{https://arxiv.org/abs/2502.17956}
\showURL{%
\tempurl}


\bibitem[Rubei et~al\mbox{.}(2025)]%
        {rubei2025prompt}
\bibfield{author}{\bibinfo{person}{Riccardo Rubei}, \bibinfo{person}{Aicha Moussaid}, \bibinfo{person}{Claudio~Di Sipio}, {and} \bibinfo{person}{Davide~Di Ruscio}.} \bibinfo{year}{2025}\natexlab{}.
\newblock \showarticletitle{Prompt Engineering and Its Implications on the Energy Consumption of Large Language Models}. In \bibinfo{booktitle}{\emph{Proceedings of the IEEE/ACM 9th International Workshop on Green and Sustainable Software (GREENS)}}. \bibinfo{pages}{1--6}.
\newblock
\href{https://doi.org/10.1109/GREENS66463.2025.00014}{doi:\nolinkurl{10.1109/GREENS66463.2025.00014}}


\bibitem[Russel et~al\mbox{.}(2024)]%
        {ai_tools_sustainability_ssrn}
\bibfield{author}{\bibinfo{person}{Asif~Hassan Russel}, \bibinfo{person}{Md~Sanaul Haque}, \bibinfo{person}{Hatef Shamshiri}, {and} \bibinfo{person}{Jari Porras}.} \bibinfo{year}{2024}\natexlab{}.
\newblock \showarticletitle{Impact of Incorporating AI Tools on Software Sustainability among Software Developers}.
\newblock \bibinfo{journal}{\emph{SSRN Electronic Journal}} (\bibinfo{year}{2024}).
\newblock
\urldef\tempurl%
\url{https://ssrn.com/abstract=5519476}
\showURL{%
\tempurl}
\newblock
\shownote{Preprint}.


\bibitem[Schaefer et~al\mbox{.}(2024)]%
        {schaefer2024_llm_unit_tests}
\bibfield{author}{\bibinfo{person}{Max Schaefer}, \bibinfo{person}{Sarah Nadi}, \bibinfo{person}{Aryaz Eghbali}, {and} \bibinfo{person}{Frank Tip}.} \bibinfo{year}{2024}\natexlab{}.
\newblock \showarticletitle{An Empirical Evaluation of Using Large Language Models for Automated Unit Test Generation}.
\newblock \bibinfo{journal}{\emph{IEEE Transactions on Software Engineering}} (\bibinfo{year}{2024}).
\newblock
\href{https://doi.org/10.1109/TSE.2023.3334955}{doi:\nolinkurl{10.1109/TSE.2023.3334955}}


\bibitem[Shi et~al\mbox{.}(2025)]%
        {shi2025_green_llm4se}
\bibfield{author}{\bibinfo{person}{Jieke Shi}, \bibinfo{person}{Zhou Yang}, {and} \bibinfo{person}{David Lo}.} \bibinfo{year}{2025}\natexlab{}.
\newblock \showarticletitle{Efficient and Green Large Language Models for Software Engineering: Vision and the Road Ahead}.
\newblock \bibinfo{journal}{\emph{ACM Transactions on Software Engineering and Methodology}} (\bibinfo{year}{2025}).
\newblock
\href{https://doi.org/10.1145/3708525}{doi:\nolinkurl{10.1145/3708525}}


\bibitem[Tang et~al\mbox{.}(2025)]%
        {tang2025fewshot}
\bibfield{author}{\bibinfo{person}{Yongjian Tang}, \bibinfo{person}{Doruk Tuncel}, \bibinfo{person}{Christian Koerner}, {and} \bibinfo{person}{Thomas Runkler}.} \bibinfo{year}{2025}\natexlab{}.
\newblock \showarticletitle{The Few-shot Dilemma: Over-prompting Large Language Models}.
\newblock \bibinfo{journal}{\emph{arXiv preprint}}  \bibinfo{volume}{arXiv:2509.13196} (\bibinfo{year}{2025}).
\newblock
\urldef\tempurl%
\url{https://arxiv.org/abs/2509.13196}
\showURL{%
\tempurl}


\bibitem[Verdecchia et~al\mbox{.}(2021a)]%
        {verdecchia_energy_testing}
\bibfield{author}{\bibinfo{person}{Roberto Verdecchia}, \bibinfo{person}{Emilio Cruciani}, \bibinfo{person}{Antonia Bertolino}, {and} \bibinfo{person}{Breno Miranda}.} \bibinfo{year}{2021}\natexlab{a}.
\newblock \showarticletitle{Energy-Aware Software Testing}. In \bibinfo{booktitle}{\emph{Proceedings of the IEEE/ACM International Conference on Software Engineering Workshops}}. \bibinfo{pages}{1--5}.
\newblock


\bibitem[Verdecchia et~al\mbox{.}(2021b)]%
        {green_it_software_2021}
\bibfield{author}{\bibinfo{person}{Roberto Verdecchia}, \bibinfo{person}{Patricia Lago}, \bibinfo{person}{Christof Ebert}, {and} \bibinfo{person}{Carol De~Vries}.} \bibinfo{year}{2021}\natexlab{b}.
\newblock \showarticletitle{Green IT and Green Software}.
\newblock \bibinfo{journal}{\emph{IEEE Software}} \bibinfo{volume}{38}, \bibinfo{number}{6} (\bibinfo{year}{2021}), \bibinfo{pages}{7--15}.
\newblock
\href{https://doi.org/10.1109/MS.2021.3102254}{doi:\nolinkurl{10.1109/MS.2021.3102254}}


\bibitem[Wang et~al\mbox{.}(2024)]%
        {wang2024_llm_testing_survey}
\bibfield{author}{\bibinfo{person}{Junjie Wang}, \bibinfo{person}{Yuchao Huang}, \bibinfo{person}{Chunyang Chen}, \bibinfo{person}{Zhe Liu}, \bibinfo{person}{Song Wang}, {and} \bibinfo{person}{Qing Wang}.} \bibinfo{year}{2024}\natexlab{}.
\newblock \showarticletitle{Software Testing with Large Language Models: Survey, Landscape, and Vision}.
\newblock \bibinfo{journal}{\emph{IEEE Transactions on Software Engineering}} (\bibinfo{year}{2024}).
\newblock
\href{https://doi.org/10.1109/TSE.2024.3368208}{doi:\nolinkurl{10.1109/TSE.2024.3368208}}


\bibitem[Wang et~al\mbox{.}(2023)]%
        {wang2023selfconsistency}
\bibfield{author}{\bibinfo{person}{Xuezhi Wang}, \bibinfo{person}{Jason Wei}, \bibinfo{person}{Dale Schuurmans}, \bibinfo{person}{Quoc Le}, \bibinfo{person}{Ed~H. Chi}, \bibinfo{person}{Sharan Narang}, \bibinfo{person}{Aakanksha Chowdhery}, {and} \bibinfo{person}{Denny Zhou}.} \bibinfo{year}{2023}\natexlab{}.
\newblock \showarticletitle{Self-Consistency Improves Chain of Thought Reasoning in Language Models}. In \bibinfo{booktitle}{\emph{Proceedings of the International Conference on Learning Representations}}.
\newblock
\urldef\tempurl%
\url{https://arxiv.org/abs/2203.11171}
\showURL{%
\tempurl}


\bibitem[Wei et~al\mbox{.}(2022)]%
        {wei2022chain}
\bibfield{author}{\bibinfo{person}{Jason Wei}, \bibinfo{person}{Xuezhi Wang}, \bibinfo{person}{Dale Schuurmans}, \bibinfo{person}{Maarten Bosma}, \bibinfo{person}{Brian Ichter}, \bibinfo{person}{Fei Xia}, \bibinfo{person}{Ed~H. Chi}, \bibinfo{person}{Quoc~V. Le}, {and} \bibinfo{person}{Denny Zhou}.} \bibinfo{year}{2022}\natexlab{}.
\newblock \showarticletitle{Chain-of-Thought Prompting Elicits Reasoning in Large Language Models}. In \bibinfo{booktitle}{\emph{Advances in Neural Information Processing Systems}}.
\newblock
\urldef\tempurl%
\url{https://arxiv.org/abs/2201.11903}
\showURL{%
\tempurl}


\bibitem[Yao et~al\mbox{.}(2023)]%
        {react2023}
\bibfield{author}{\bibinfo{person}{Shunyu Yao}, \bibinfo{person}{Jeffrey Zhao}, \bibinfo{person}{Dian Yu}, \bibinfo{person}{Nan Du}, \bibinfo{person}{Izhak Shafran}, \bibinfo{person}{Karthik Narasimhan}, {and} \bibinfo{person}{Yuan Cao}.} \bibinfo{year}{2023}\natexlab{}.
\newblock \showarticletitle{ReAct: Synergizing Reasoning and Acting in Language Models}. In \bibinfo{booktitle}{\emph{Proceedings of the International Conference on Learning Representations (ICLR)}}.
\newblock
\urldef\tempurl%
\url{https://arxiv.org/abs/2210.03629}
\showURL{%
\tempurl}


\bibitem[Zaidman(2024)]%
        {zaidman_ast2024}
\bibfield{author}{\bibinfo{person}{Andy Zaidman}.} \bibinfo{year}{2024}\natexlab{}.
\newblock \showarticletitle{An Inconvenient Truth in Software Engineering? The Environmental Impact of Testing Open Source Java Projects}. In \bibinfo{booktitle}{\emph{Proceedings of the 5th ACM/IEEE International Conference on Automation of Software Test (AST)}}. \bibinfo{pages}{1--5}.
\newblock
\href{https://doi.org/10.1145/3644032.3644461}{doi:\nolinkurl{10.1145/3644032.3644461}}


\bibitem[Zhou et~al\mbox{.}(2023)]%
        {zhou2023least}
\bibfield{author}{\bibinfo{person}{Denny Zhou}, \bibinfo{person}{Nathanael Scharli}, \bibinfo{person}{Le Hou}, \bibinfo{person}{Jason Wei}, \bibinfo{person}{Nathan Scales}, \bibinfo{person}{Xuezhi Wang}, \bibinfo{person}{Dale Schuurmans}, \bibinfo{person}{Claire Cui}, \bibinfo{person}{Olivier Bousquet}, \bibinfo{person}{Quoc Le}, {and} \bibinfo{person}{Ed Chi}.} \bibinfo{year}{2023}\natexlab{}.
\newblock \showarticletitle{Least-to-Most Prompting Enables Complex Reasoning in Large Language Models}. In \bibinfo{booktitle}{\emph{Proceedings of the International Conference on Learning Representations (ICLR)}}.
\newblock
\urldef\tempurl%
\url{https://arxiv.org/abs/2205.10625}
\showURL{%
\tempurl}


\end{thebibliography}
\end{document}